\documentclass{PoS}

\usepackage{amsmath,amssymb,mathrsfs,dsfont,
subcaption,mathtools,verbatim}

\usepackage{scalerel}
\usepackage{feynmp-auto}
\newcommand{\dfn}{\vcentcolon=}

\fmfcmd{%
vardef cross_bar (expr p, len, ang) = 
((-len/2,0)--(len/2,0)) rotated (ang + angle direction length(p)/2 of p) 
shifted point length(p)/2 of p
enddef;
style_def crossed expr p =
cdraw p;
ccutdraw cross_bar (p, 5mm,  45);
ccutdraw cross_bar (p, 5mm, -45)
enddef;}

\title{Tadpoles, Cephalopods, and `Complete Normal Ordering'}

\ShortTitle{Complete Normal Ordering}

\author{\speaker{Dimitri Skliros}%
        \thanks{This work summarises part of a more detailed paper to appear by the present author in collaboration with John Ellis and Nikolaos E.~Mavromatos.}\\
       Theoretical Particle Physics and Cosmology Group, Department of Physics,\\ 
       King's College London, Strand, London WC2R 2LS, U.K\\
       E-mail: \email{dimitris.skliros@kcl.ac.uk}}

\abstract{We describe how to cancel (when this is desirable) all tadpole and more generally all cephalopod Feynman diagrams in generic interacting scalar field theories to all orders in perturbation theory. This cancelation reduces the number of Feynman diagrams at a given loop order by a factor of two or more, and is accomplished by introducing the notion of `complete normal ordering' (an extension of the standard field theory definition of normal ordering) which when applied to the bare action of the theories of interest results in tadpole- and cephalopod-free Greens functions.}

\FullConference{18th International Conference From the Planck Scale to the Electroweak Scale \\
		 25-29 May 2015\\
		 Ioannina, Greece }

\begin{document}

\section{Overview}
Consistent string theories are normally phrased in terms of two-dimensional (super)conformal field theories. More generally, it might be suspected that \cite{Polchinski_v12} the known string theories should be thought of as (super)conformal fixed points of all two-dimensional (possibly supersymmetric) quantum field theories, and this motivates the study of generic two-dimensional quantum field theories, their associated renormalisation group (RG) flows, fixed points, etc., whereby the various RG flows connect the various 2d superconformal field theories. On a parallel note, string theories in non-trivial backgrounds are usually phrased in terms of non-linear sigma models. These are 2-dimensional non-linear field theories with derivative interactions, whose local coupling constants correspond to background quantities such as the spacetime metric, dilaton, etc. In all these cases, a starting point is to study the RG flow of these coupling constants, and so one needs to understand how to renormalise the theory. 

Although there is a vast literature on the renormalisation of non-linear sigma models, see e.g.~\cite{nlsm} 
and references therein, there are numerous subtleties that have not yet (to the best of my knowledge) been worked out fully, such as the non-linear renormalisation of quantum fields \cite{HowePapadopoulosStelle88}, the role of string loops (see e.g.~\cite{RGstringloops}), and the potential presence of moduli or zero modes that parametrise the classical background around which the background field expansion is carried out, as these presumably should also be integrated out in order to derive the effective worldsheet action from which the beta functions follow. Looking ahead, having derived consistency conditions for allowed string backgrounds, one would like to go on to quantise strings in such backgrounds, construct vertex operators \cite{CallanGan86}, correlation functions, and this is also likely to be a fundamental step in studies of strings in the early universe and in the context of black holes. 

In what follows we focus on a tiny aspect of this ambitious program. To place things into context, a fundamental tool in generic quantum field theories is the notion of `normal ordering', $\mathcal{O}(\phi)\rightarrow \,\,:\!\mathcal{O}(\phi)\!:\,$, which becomes indispensable, e.g., when defining operators at coincident points or in the evaluation of correlation functions using Wick's theorem. For instance, normal-ordered operators have the useful property that their expectation values in the \emph{free} theory vanish: $\langle:\!\!\mathcal{O}(\phi)\!\!:\rangle_0=0$. There are various guises of this notion \cite{Polchinski_v12}, such as creation-annihilation operator normal ordering, conformal normal ordering and functional integral normal ordering, etc. These are often interrelated. A concise definition can be given at the level of the functional integral:\footnote{This is equivalent to the usual definition \cite{Polchinski_v12} $:\!\mathcal{O}(\phi)\!:\,\,=\exp\big(-\frac{1}{2}\int_{z}\int_{w}\mathcal{G}(z,w)\frac{\delta}{\delta \phi(z)}\frac{\delta}{\delta \phi(w)}\big)\,\mathcal{O}(\phi)$; 
however the above expression will be more useful in what follows for reasons that will become clear.}
\begin{equation}\label{eq:NO}
: \!\mathcal{O}(\phi)\!:\,\,=\mathcal{O}(\delta_X)\,e^{-W_0(X)+\int X\phi}\big|_{X=0},
\end{equation}
where $W_0(X)$ is the renormalised generating function of the \emph{free} Feynman propagator of the theory, call it $\mathcal{G}(z,w)$, (obtained from (\ref{eq:W(J)}) by setting interactions to zero) and $\delta_X$ a functional derivative with respect to the (unphysical) source $X$. 
And so normal ordering is carried out at the level of the free theory. In the interaction picture of quantum field theory this becomes a useful concept: for example, normal ordering the action results in correlation functions that are free from Feynman diagrams with internal lines that begin and end on the same internal vertex. In this manner, certain (but \emph{not} all) tadpole diagrams in $\phi^3$ scalar field theory are cancelled, as is the one-loop two-point amplitude in $\phi^4$, to name a couple of examples. Normal ordering is certainly also an indispensable tool in string theory where, e.g., it is particularly efficient to represent external states by normal ordered vertex operators, or to define the quantum stress-energy tensor. Of course, normal ordering does not change the quantum field theory in any observable way, given that it can always be undone by a particular choice of counter terms. Normal ordering is also not unique: one can obtain a different prescription by replacing $\mathcal{G}(z,w)$ in (\ref{eq:NO}) by another two-point function $\mathcal{G}'(z,w)=\mathcal{G}(z,w)+\Delta(z,w)$, and one has to make a particular choice in order for the aforementioned tadpole diagrams to cancel.

The fact that \emph{certain} tadpole diagrams are cancelled by normal ordering the action (with often an infinite number of tadpole diagrams remaining) is curious, and the fact that expectation values of normal ordered operators, $\langle:\!\!\mathcal{O}(\phi)\!\!:\rangle$, only vanish in the \emph{free} theory suggests that the definition of normal ordering can be improved. It would be extremely valuable if we could construct a new form of normal ordering, let us call it `complete normal ordering', that ensures \emph{all} (either massless or massive) tadpoles are cancelled to all orders in perturbation theory, and that the expectation value of a `complete normal ordered operator' computed in the full interacting theory vanishes identically.\footnote{This is to be contrasted with `normal products' or `composite operators' where one requires that correlation functions involving generic insertions of such operators and elementary fields are well-defined \cite{Zimmermann73}.}

In what follows we do precisely this: we introduce a map $\mathcal{O}(\phi)\rightarrow \,*\mathcal{O}(\phi)*$ called `\emph{complete normal ordering}' that: 
\begin{itemize}
\item[(i)] ensures that a large class of Feynman diagrams (that we call `cephalopod' diagrams\footnote{\label{foot}The terminology `cephalopod' (which literally means `head-feet') is borrowed from marine biology where cephalopods are marine animals characterised by bilateral body symmetry, they have a prominent head and a set of tentacles. The use here is meant to be suggestive: the 1PI version of a `cephalopod' Feynman graph has an arbitrary number (0,1,2,..) of ``legs'' and an arbitrary number (1,2,3,..) of ``heads''. There is no restriction on the number of loops the head is composed of, other than that the ``neck'' joining the head(s) and leg(s) is represented by a single vertex. This class of diagrams \emph{includes all} tadpole diagrams but is generically a much larger class of diagrams. The precise definition and examples are given on p.~\pageref{cephalopod-dfn}.}) are cancelled from all correlation functions to all orders in perturbation theory
\item[(ii)] provides an explicit expression for the counter terms that accomplish this cancellation
\end{itemize}
Complete normal ordering is (as the name suggests) more complete than the usual notion of `normal ordering' \cite{Polchinski_v12}. In the latter case one subtracts all self contractions from a given operator using Wick's theorem (and the free two-point function), and when applied to, say, the Lagrangian under consideration, $\mathcal{L}(\phi)\rightarrow \,\,:\!\mathcal{L}(\phi)\!:\,$, this cancels all Feynman diagrams with internal lines that begin and end on the same vertex.  `Complete normal ordering' generalises these notions by instead subtracting \emph{all} self contractions using a generalisation of Wick's theorem, whereby the subtractions are carried out with the full renormalised $N$-point Greens functions. (Therefore, there does not exist a prescription (a choice of $\Delta(z,w)$, see above) that maps $:\!\mathcal{O}\!:\rightarrow *\mathcal{O}*$, although the inverse map, $*\mathcal{O}*\rightarrow :\!\mathcal{O}\!:$, does always exist.) A concise definition of complete normal ordered operators is:
\begin{equation}\label{eq:CNO}
* \,\mathcal{O}(\phi)\,*=\mathcal{O}(\delta_X)\,e^{-W(X)+W(0)+\int X\phi}\big|_{X=0},
\end{equation}
where $W(X)$ is the renormalised generating function of all connected Greens functions in the theory of interest, and $X$ and $\phi$ a renormalised source and field respectively. 

When applied to the Lagrangian, $\mathcal{L}(\phi)\rightarrow \,\,*\mathcal{L}(\phi)*\,$, all diagrams cancelled by normal ordering are also cancelled in complete normal ordering, but in addition all tadpoles and all cephalopods more generally are also cancelled. This is particularly useful in explicit loop computations: for example, in the context of Liouville field theory, the presence of numerous tadpoles is one of the main stumbling blocks for going beyond low orders in loop perturbation theory \cite{Zams}: 
\begin{quote}
`\emph{Here we will not develop further the loop perturbation theory for LFT [Liouville Field Theory] on the Lobachevskiy plane. To go at higher loop diagrammatic calculations it is worth first to improve the technique to better handle the tadpole diagrams (which become rather numerous at higher orders)}\dots'
\end{quote}
Complete normal ordering cancels \emph{all} tadpole diagrams. Furthermore, complete normal ordered operators also have vanishing expectation values in the full \emph{interacting} theory (unless there is a physical source): $\langle*\mathcal{O}(\phi)*\rangle=0$. 

Complete normal ordering works in any number of spacetime dimensions and is to a large extent independent of the background spacetime on which the field theory is formulated (under the usual assumptions, such as the requirement of global hyperbolicity).  In certain cases, for example two-dimensional quantum field theories without derivative interactions, simple power counting suggests that complete normal ordering automatically renders all correlation functions of elementary fields UV finite while providing an explicit expression for the counter terms that are ultimately responsible for the associated subtractions. We study the case of a single scalar field for simplicity, but expect the basic formalism to be much more general.

%%%%%%%%%%%%%%%%%%%%%%%%%%%%%%%%%%%%%%%%%%%%%%%%%%%%%%%%%%%%%%%%%%%%%%%
\section{Tadpoles, Cephalopods, and `Complete Normal Ordering'}

Suppose we are given an interacting quantum field theory of a single scalar field, $\phi$, defined on some fixed globally hyperbolic spacetime background of integer spacetime dimension $n$ and metric $ds^2=g_{\alpha\beta}dz^{\alpha}dz^{\beta}$, continued to Euclidean space. Denote by\footnote{A diffeomorphism-invariant measure, $d\mu_g=d^dz\sqrt{g}$, is always implied when not displayed explicitly.} $I_B(\phi_B)=\int \mathcal{L}_B(\phi_B)$ its `bare' action, with bare Lagrangian $\mathcal{L}_B(\phi_B)$. In the spirit of dimensional regularisation we have continued the spacetime dimension \cite{'tHooft73} to $d=n-2\epsilon$, with $\epsilon$ the dimensional regularisation parameter, and we denote by $\mu$ the associated mass scale with respect to which renormalised couplings run. The scalar field has mass dimension $\frac{d}{2}-1$. Let us also denote the associated generating function of connected Greens functions by $W(J)$, with $J$ the associated renormalised source. (Note that bare and renormalised generating functions are equal.) Renormalised connected $N$-point Greens functions, $G_N(z_1,\dots,z_N)$, are then defined by a formal expansion in powers of the source:
\begin{equation}\label{eq:W(J)}
W(J)=\sum_{N=0}^{\infty}\frac{1}{N!}\int_1\dots\int_NG_N(z_1,\dots,z_N)J(z_1)\dots J(z_N),
\end{equation}
and extracted from $W(J)$ by repeated functional differentiation\footnote{We denote functional derivatives with respect to a generic field $X(z)$ by $\delta_{X(z)}\dfn \frac{1}{\sqrt{g}}\frac{\delta }{\delta X(z)}$.}. The generating function is in turn related to the bare action via $e^{W(J)}=\int \mathcal{D}\phi_Be^{-I_B(\phi_B)}$, where we often absorb the (possibly physical) source into the action: $-I_B(\phi_B)\supset \int J_B\phi_B$, with the renormalised and bare sources $J$ and $J_B$ respectively related below. Correlation functions of some collection of renormalised fields $\mathcal{O}(\phi)$ are always defined with respect to this path integral, $\langle \mathcal{O}(\phi)\rangle\dfn \mathcal{O}(\delta_J)e^{W(J)}$, with generic normalisation $\langle1\rangle=e^{W(0)}$. (Of course, when the source $J$ is unphysical we set it to zero after evaluating the functional derivatives.)

This theory may or may not be perturbatively renormalisable, and what follows will not depend on this. Let us make explicit the wavefunction renormalisation $Z=1+\delta Z$ (with $\phi_B=Z\phi$ and $\phi$ the renormalised field) and various bare couplings $\{g_N^B|N=0,1,2,\dots\}$ (associated to mass-dimension $K(N)\equiv N(\frac{d}{2}-1)$ operators\footnote{The example we have in mind is in the absence of derivative interactions, but we expect the results below to hold also for derivative interactions when the corresponding bare action has the counter terms generated by complete normal ordering. Derivative interactions will be discussed in greater detail in \cite{EllisMavromatosSkliros15}.}) 
that appear:
\begin{equation}\label{eq:Notation}
\begin{aligned}
\mathcal{L}_B(\phi_B)&=\mathcal{L}_B(g_1^B,g_2^B,\dots;Z)\\
&=\mathcal{L}_B\Big(\mu^{-\epsilon}(g_1+\delta g_1)Z^{-1/2},(g_2+\delta g_2)Z^{-1},\dots,\mu^{(N-2)\epsilon}(g_N+\delta g_N)Z^{-N/2},\dots;Z\Big).
\end{aligned}
\end{equation}
In the second line the quantities $g_N$ are renormalised couplings and the $\delta g_N$ the corresponding counter terms. 
By dimensional analysis (note that $K(1)=\frac{d}{2}-1$ is the mass dimension of the scalar field), the renormalised parameter $g_1$ will correspond to a renormalised source, call it $g_1\equiv-J$, related to the bare source via, $J_B=\mu^{-\epsilon}(J+\delta J)/Z^{1/2}$, as implied in (\ref{eq:Notation}); $g_2$ will have the interpretation of a renormalised mass, $g_2=m^2$, with corresponding counter term $\delta g_2=\delta m^2$, $g_3$ may correspond to some cubic interaction ``coupling constant'' (which need not be a true constant), and so on. (Generically, the mass dimensions of renormalised couplings, $n-N(\frac{n}{2}-1)$, are independent of $\epsilon$.)

\vspace{0.2cm}
The counter terms, $\delta g_N$, $\delta Z$, will at a generic loop order get contributions from a large number of Feynman diagrams. 
In what follows we will explain how to remove a large subclass of these diagrams from the generating function of connected Greens functions, $W(J)$, (all `cephalopod' Feynman diagrams in particular, see below) and extract the associated counter terms that accomplish this cancellation:
\begin{itemize}
\item \underline{`\emph{Cephalopod' Feynman diagrams}}\label{cephalopod-dfn}:\footnote{Recall the second footnote on p.~\pageref{foot}.} these are connected diagrams that can be disconnected into two pieces by cutting one internal \emph{vertex}\footnote{By `cutting a diagram across a vertex' it is meant that if an internal $N$-point vertex is cut, the resulting two subdiagrams will contain an (incomplete) $N-m$- and an (incomplete) $m$-point vertex respectively, the other constituents of the graphs remaining unchanged.}  but with either one or both resulting pieces free from external lines; examples are: 
$$
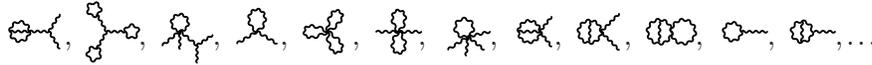
\begin{gathered}
	\begin{fmffile}{wg2L-3loop2pt-1PR-}
		\begin{fmfgraph}(16,16)
			\fmfset{dash_len}{1.2mm}
			\fmfset{wiggly_len}{1.1mm} \fmfset{dot_len}{0.5mm}
			\fmfpen{0.25mm}
			\fmfleft{i}
			\fmfright{o}
			\fmf{phantom,tension=5}{i,v1}
			\fmf{wiggly,fore=black,tension=2.5}{v2,o}
			\fmf{wiggly,fore=black,left,tension=0.5}{v1,v2,v1}
			\fmf{wiggly,fore=black}{v1,v2}
			\fmffreeze
			\fmfforce{(1.1w,0.5h)}{o}
			\fmffreeze
			\fmfright{n,m}
			\fmf{wiggly,fore=black,tension=1}{o,n}
			\fmf{wiggly,fore=black,tension=1}{o,m}
			\fmfforce{(1.35w,0.9h)}{n}
			\fmfforce{(1.35w,0.1h)}{m}
		\end{fmfgraph}
	\end{fmffile}
\end{gathered}
\,\,\,\,\,\,, 
\!\! \!\!\!\!\!\!
\quad\,\,
\begin{gathered}
	\begin{fmffile}{wg4-3tadpole-}
		\begin{fmfgraph}(14,14)
			\fmfset{dash_len}{1.2mm}
			\fmfset{wiggly_len}{1.1mm} \fmfset{dot_len}{0.5mm}
			\fmfpen{0.25mm}
			\fmfsurround{u1,u2,u3}
			\fmf{wiggly,fore=black,tension=1}{u1,v}
			\fmf{wiggly,fore=black,tension=1}{u2,v}
			\fmf{wiggly,fore=black,tension=1}{u3,v}
			\fmf{wiggly,fore=black,tension=1,left}{u1,u1}
			\fmf{wiggly,fore=black,tension=1,left}{u2,u2}
			\fmf{wiggly,fore=black,tension=1,left}{u3,u3}
		\end{fmfgraph}
	\end{fmffile}
\end{gathered}\,\,\,\,,\,\,
\,
\begin{gathered}
	\begin{fmffile}{wgkappa-4ptself1PR-}
		\begin{fmfgraph}(13,13)
			\fmfset{dash_len}{1.2mm}
			\fmfset{wiggly_len}{1.1mm} \fmfset{dot_len}{0.5mm}
			\fmfpen{0.25mm}
			\fmftop{t}
			\fmfbottom{a,b,c}
			\fmf{wiggly,fore=black,tension=1}{a,v}
			\fmf{wiggly,fore=black,tension=1}{b,v}
			\fmf{wiggly,fore=black,tension=1}{c,v}
			\fmf{wiggly,fore=black,tension=1.4,left}{v,t,v}
			\fmffreeze
			\fmfbottom{x,z}
			\fmf{wiggly,fore=black}{c,x}
			\fmf{wiggly,fore=black}{c,z}
			\fmfforce{(w,-.4h)}{x}
			\fmfforce{(1.45w,.4h)}{z}
		\end{fmfgraph}
	\end{fmffile}
\end{gathered}\,\,
\,\,\,, \,\,\,
\begin{gathered}
	\begin{fmffile}{wlambdaself-}
		\begin{fmfgraph}(15,15)
			\fmfset{dash_len}{1.2mm}
			\fmfset{wiggly_len}{1.1mm} \fmfset{dot_len}{0.5mm}
			\fmfpen{0.25mm}
			\fmftop{s}
			\fmfleft{a}
			\fmfright{b}
			\fmf{wiggly,fore=black}{a,v}
			\fmf{wiggly,fore=black}{b,v}
			\fmf{wiggly,fore=black,right,tension=.7}{v,v}
			\fmffreeze
			\fmfforce{(0w,0.2h)}{a}
			\fmfforce{(w,0.2h)}{b}
		\end{fmfgraph}
	\end{fmffile}
\end{gathered}
\,, 
\,\,\,
\begin{gathered}
	\begin{fmffile}{wgamma-vacuum-1PI-}
		\begin{fmfgraph}(18,18)
			\fmfset{dash_len}{1.2mm}
			\fmfset{wiggly_len}{1.1mm} \fmfset{dot_len}{0.5mm}
			\fmfpen{0.25mm}
			\fmfsurroundn{x}{3}
			\fmf{phantom,fore=black}{x1,v}
			\fmf{phantom,fore=black}{x2,v}
			\fmf{phantom,fore=black}{x3,v}
			\fmf{wiggly,fore=black,tension=0.7}{v,v}
			\fmf{wiggly,fore=black,tension=0.7,right}{v,v}
			\fmf{wiggly,fore=black,tension=0.7,left}{v,v}
		\end{fmfgraph}
	\end{fmffile}
\end{gathered}, 
\,\,\,
\begin{gathered}
	\begin{fmffile}{wgamma-2pt1loop-1PI-}
		\begin{fmfgraph}(17,17)
			\fmfset{dash_len}{1.2mm}
			\fmfset{wiggly_len}{1.1mm} \fmfset{dot_len}{0.5mm}
			\fmfpen{0.25mm}
			\fmfleft{i}
			\fmfright{o}
			\fmf{wiggly,fore=black,tension=0.7}{i,v,v,o}
			\fmf{wiggly,fore=black,left=90,tension=0.7}{v,v}
		\end{fmfgraph}
	\end{fmffile}
\end{gathered}\!, 
\,\,\,
\begin{gathered}
	\begin{fmffile}{wgamma-4pt1loop-1PI-}
		\begin{fmfgraph}(18,18)
			\fmfset{dash_len}{1.2mm}
			\fmfset{wiggly_len}{1.1mm} \fmfset{dot_len}{0.5mm}
			\fmfpen{0.25mm}
			\fmfsurroundn{x}{8}
			\fmf{phantom,fore=black}{x1,c,x5}
			\fmf{phantom,fore=black}{x2,c,x6}
			\fmf{phantom,fore=black}{x3,c,x7}
			\fmf{phantom,fore=black}{x4,c,x8}
			\fmf{wiggly,fore=black}{x1,c}
			\fmf{wiggly,fore=black}{x8,c}
			\fmf{wiggly,fore=black}{x7,c}
			\fmf{wiggly,fore=black}{x6,c}
			\fmfi{wiggly,fore=black}{fullcircle scaled .38w shifted (0.46w,.58h)}	
		\end{fmfgraph}
	\end{fmffile}
\end{gathered},
\,\,
\begin{gathered}
	\begin{fmffile}{wgkappa-bubble-2pt-1PI-}
		\begin{fmfgraph}(14,14)
			\fmfset{dash_len}{1.2mm}
			\fmfset{wiggly_len}{1.1mm} \fmfset{dot_len}{0.5mm}
			\fmfpen{0.25mm}
			\fmfleft{i}
			\fmfright{o}
			\fmf{phantom,tension=5}{i,v1}
			\fmf{phantom,tension=5}{v2,o}
			\fmf{wiggly,fore=black,left,tension=0.4}{v1,v2,v1}
			\fmf{wiggly,fore=black}{v1,v2}
			\fmffreeze
			\fmfright{o1,o2}
			\fmf{wiggly,fore=black,tension=1}{v2,o1}
			\fmf{wiggly,fore=black,tension=1}{v2,o2}
			\fmfforce{(1.1w,0.9h)}{o1}
			\fmfforce{(1.1w,0.1h)}{o2}
		\end{fmfgraph}
	\end{fmffile}
\end{gathered}\,
\,, \,\,\,\,\,\,
\!\!\!\begin{gathered}
	\begin{fmffile}{wg2L-bubble2ptz-}
		\begin{fmfgraph}(17,17)
			\fmfset{dash_len}{1.2mm}
			\fmfset{wiggly_len}{1.1mm} \fmfset{dot_len}{0.5mm}
			\fmfpen{0.25mm}
			\fmftop{t1,t2,t3}
			\fmfbottom{b1,b2,b3}
			\fmf{phantom}{t1,v1,b1}
			\fmf{phantom}{t2,v2,b2}
			\fmf{phantom}{t3,v3,b3}
			\fmffreeze
			\fmf{wiggly,fore=black,right}{v1,v2,v1}
			\fmf{wiggly,fore=black}{v2,t3}
			\fmf{wiggly,fore=black}{v2,b3}
			\fmf{wiggly,fore=black,tension=1}{t1,b1}
			\fmfforce{(0.25w,0.7h)}{t1}
			\fmfforce{(0.25w,0.3h)}{b1}
			\fmfforce{(0.9w,0.9h)}{t3}
			\fmfforce{(0.9w,0.1h)}{b3}
		\end{fmfgraph}
	\end{fmffile}
\end{gathered}, 
\,\,\,
\begin{gathered}
	\begin{fmffile}{wg2L-bubble-}
		\begin{fmfgraph}(18,18)
			\fmfset{dash_len}{1.2mm}
			\fmfset{wiggly_len}{1.1mm} \fmfset{dot_len}{0.5mm}
			\fmfpen{0.25mm}
			\fmftop{t1,t2,t3}
			\fmfbottom{b1,b2,b3}
			\fmf{phantom}{t1,v1,b1}
			\fmf{phantom}{t2,v2,b2}
			\fmf{phantom}{t3,v3,b3}
			\fmffreeze
			\fmf{wiggly,fore=black,right}{v1,v2,v1}
			\fmf{wiggly,fore=black,right}{v2,v3,v2}
			\fmf{wiggly,fore=black,tension=1}{t1,b1}
			\fmfforce{(0.25w,0.7h)}{t1}
			\fmfforce{(0.25w,0.3h)}{b1}
		\end{fmfgraph}
	\end{fmffile}
\end{gathered},\,\,\,
\begin{gathered}
	\begin{fmffile}{wgtadpole-}
		\fmfset{dash_len}{1.2mm}
		\begin{fmfgraph}(18,18)
			\fmfset{dash_len}{1.2mm}
			\fmfset{wiggly_len}{1.1mm} \fmfset{dot_len}{0.5mm}
			\fmfpen{0.25mm}
			\fmfleft{i}
			\fmfright{o}
			\fmf{phantom,tension=5}{i,v1}
			\fmf{wiggly,fore=black,tension=0.8}{v2,o}
			\fmf{wiggly,fore=black,left,tension=0.08}{v1,v2,v1}
			\fmf{phantom}{v1,v2}
		\end{fmfgraph}
	\end{fmffile}
\end{gathered}
,  
\!\!\!\!\!\!
\,\,
\begin{gathered}
	\begin{fmffile}{wg3tadpole1-}
		\begin{fmfgraph}(25,25)
			\fmfset{dash_len}{1.2mm}
			\fmfset{wiggly_len}{1.1mm} \fmfset{dot_len}{0.5mm}
			\fmfpen{0.25mm}
			\fmftop{t}
			\fmfbottom{b}
			\fmfleft{l}
			\fmfright{r}
			\fmf{phantom,fore=black,tension=9}{t,u,v,b}
			\fmf{phantom,fore=black,tension=9}{l,s,x,r}
			\fmf{wiggly,fore=black,tension=.01,left}{u,v,u}
			\fmf{phantom,fore=black,tension=0.01}{s,x,s}
			\fmf{wiggly,fore=black,tension=1}{u,v}
			\fmf{wiggly,fore=black,tension=1}{x,r}
		\end{fmfgraph}
	\end{fmffile}
\end{gathered}
,
\dots
$$ 
There exist both one-particle irreducible (1PI) and one-particle reducible (1PR) cephalopod graphs --the first three depicted are all 1PR whereas the remaining ones are 1PI. 
A \emph{subset} of all cephalopod diagrams are the \emph{tadpole diagrams}, of which there are two types: 
\begin{itemize}
\item[(i)] \underline{\emph{1PI tadpole diagrams}}: 1PI cephalopod diagrams with a \emph{single} external line and any number of loops, such as: 
$$
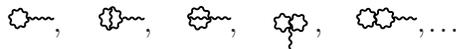
\begin{gathered}
	\begin{fmffile}{wgtadpole-}
		\fmfset{dash_len}{1.2mm}
		\begin{fmfgraph}(18,18)
			\fmfset{dash_len}{1.2mm}
			\fmfset{wiggly_len}{1.1mm} \fmfset{dot_len}{0.5mm}
			\fmfpen{0.25mm}
			\fmfleft{i}
			\fmfright{o}
			\fmf{phantom,tension=5}{i,v1}
			\fmf{wiggly,fore=black,tension=0.8}{v2,o}
			\fmf{wiggly,fore=black,left,tension=0.08}{v1,v2,v1}
			\fmf{phantom}{v1,v2}
		\end{fmfgraph}
	\end{fmffile}
\end{gathered}
,  
\!\!\!\!\!\!
\quad
\begin{gathered}
	\begin{fmffile}{wg3tadpole1-}
		\begin{fmfgraph}(25,25)
			\fmfset{dash_len}{1.2mm}
			\fmfset{wiggly_len}{1.1mm} \fmfset{dot_len}{0.5mm}
			\fmfpen{0.25mm}
			\fmftop{t}
			\fmfbottom{b}
			\fmfleft{l}
			\fmfright{r}
			\fmf{phantom,fore=black,tension=9}{t,u,v,b}
			\fmf{phantom,fore=black,tension=9}{l,s,x,r}
			\fmf{wiggly,fore=black,tension=.01,left}{u,v,u}
			\fmf{phantom,fore=black,tension=0.01}{s,x,s}
			\fmf{wiggly,fore=black,tension=1}{u,v}
			\fmf{wiggly,fore=black,tension=1}{x,r}
		\end{fmfgraph}
	\end{fmffile}
\end{gathered}
,\!\! 
\quad
\begin{gathered}
	\begin{fmffile}{wglambdatadpole-}
		\begin{fmfgraph}(17,17)
			\fmfset{dash_len}{1.2mm}
			\fmfset{wiggly_len}{1.1mm} \fmfset{dot_len}{0.5mm}
			\fmfpen{0.25mm}
			\fmfleft{i}
			\fmfright{o}
			\fmf{phantom,tension=5}{i,v1}
			\fmf{wiggly,fore=black,tension=2.5}{v2,o}
			\fmf{wiggly,fore=black,left,tension=0.5}{v1,v2,v1}
			\fmf{wiggly,fore=black}{v1,v2}
			\fmffreeze
			\fmfforce{(1.1w,0.5h)}{o}
		\end{fmfgraph}
	\end{fmffile}
\end{gathered}
\,, \!\!\!\!
\quad
\begin{gathered}
	\begin{fmffile}{wkappa-tadpole-}
		\begin{fmfgraph}(23,23)
			\fmfset{dash_len}{1.2mm}
			\fmfset{wiggly_len}{1.1mm} \fmfset{dot_len}{0.5mm}
			\fmfpen{0.25mm}
			\fmftop{t1,t2,t3}
			\fmfbottom{b1,b2,b3}
			\fmf{phantom}{t1,v1,b1}
			\fmf{phantom}{t2,v2,b2}
			\fmf{phantom}{t3,v3,b3}
%			\fmffreeze
			\fmf{wiggly,fore=black,right,tension=1}{v1,v2,v1}
			\fmf{wiggly,fore=black,right,tension=1}{v2,v3,v2}
			\fmf{wiggly,fore=black,tension=1}{v2,b2}
		\end{fmfgraph}
	\end{fmffile}
\end{gathered}\!\!
, 
\quad
\begin{gathered}
	\begin{fmffile}{wglambda-tadpole-}
		\begin{fmfgraph}(20,20)
			\fmfset{dash_len}{1.2mm}
			\fmfset{wiggly_len}{1.1mm} \fmfset{dot_len}{0.5mm}
			\fmfpen{0.25mm}
			\fmftop{t1,t2,t3,t4}
        			\fmfbottom{b1,b2,b3,b4}
        			\fmf{phantom}{t1,v1,b1}
        			\fmf{phantom}{t2,v2,b2}
			\fmf{phantom}{t3,v3,b3}
			\fmf{phantom}{t4,v4,b4}
        			\fmffreeze
			\fmf{wiggly,fore=black,right,tension=0.7}{v1,v2,v1}
        			\fmf{wiggly,fore=black,right,tension=0.7}{v2,v3,v2}
        			\fmf{wiggly,fore=black,tension=3}{v3,v4}
			\fmffreeze
			\fmfforce{(1.1w,0.5h)}{v4}
		\end{fmfgraph}
	\end{fmffile}
\end{gathered}\,\,,
\dots
$$
\item[(ii)] \underline{\emph{1PR tadpole diagrams}}: 1PR cephalopod diagrams (that can be disconnected into two pieces by cutting one internal line) but with \emph{either} one or both resulting pieces having a single external line, such as 
\!
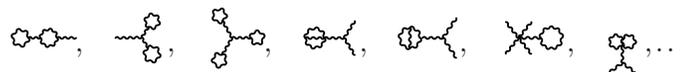
$$
\begin{gathered}
	\begin{fmffile}{wg3-tadpole2-}
		\begin{fmfgraph}(20,20)
			\fmfset{dash_len}{1.2mm}
			\fmfset{wiggly_len}{1.1mm} \fmfset{dot_len}{0.5mm}
			\fmfpen{0.25mm}
			\fmfleft{i}
			\fmfright{o}
			\fmf{phantom,tension=5}{i,v1}
			\fmf{wiggly,fore=black,tension=.04}{v2,o}
			\fmf{wiggly,fore=black,left,tension=0.01}{v1,v3,v1}
			\fmf{wiggly,fore=black,right,tension=0.01}{v2,v4,v2}
			\fmf{wiggly,fore=black,tension=0.03}{v3,v4}
			\fmffreeze
			\fmfforce{(1.2w,0.5h)}{o}
			\fmfforce{(.3w,0.5h)}{v3}
		\end{fmfgraph}
	\end{fmffile}
\end{gathered} \,\,\,,\!\! \!\!
 \,\,\,
\quad
\begin{gathered}
	\begin{fmffile}{wg3-2tadpoles-}
		\begin{fmfgraph}(17,17)
			\fmfset{dash_len}{1.2mm}
			\fmfset{wiggly_len}{1.1mm} \fmfset{dot_len}{0.5mm}
			\fmfpen{0.25mm}
			\fmfleft{i}
			\fmfright{o1,o2}
			\fmf{wiggly,fore=black,tension=1}{i,v1}
			\fmf{phantom,tension=1}{v1,u1,01}
			\fmf{phantom,tension=1}{v1,u2,o2}
			\fmf{wiggly,fore=black,tension=0.8,left}{u1,o1,u1}
			\fmf{wiggly,fore=black,tension=0.4,right}{u2,o2,u2}
			\fmf{wiggly,fore=black,tension=1}{v1,u1}
			\fmf{wiggly,fore=black,tension=1}{v1,u2}
		\end{fmfgraph}
	\end{fmffile}
\end{gathered}
\,\,,\, 
\quad
\begin{gathered}
	\begin{fmffile}{wg4-3tadpole-}
		\begin{fmfgraph}(14,14)
			\fmfset{dash_len}{1.2mm}
			\fmfset{wiggly_len}{1.1mm} \fmfset{dot_len}{0.5mm}
			\fmfpen{0.25mm}
			\fmfsurround{u1,u2,u3}
			\fmf{wiggly,fore=black,tension=1}{u1,v}
			\fmf{wiggly,fore=black,tension=1}{u2,v}
			\fmf{wiggly,fore=black,tension=1}{u3,v}
			\fmf{wiggly,fore=black,tension=1,left}{u1,u1}
			\fmf{wiggly,fore=black,tension=1,left}{u2,u2}
			\fmf{wiggly,fore=black,tension=1,left}{u3,u3}
		\end{fmfgraph}
	\end{fmffile}
\end{gathered}\,\,\,\,
, \!\!\!
\quad
\begin{gathered}
	\begin{fmffile}{wg2L-3loop2pt-1PR-}
		\begin{fmfgraph}(16,16)
			\fmfset{dash_len}{1.2mm}
			\fmfset{wiggly_len}{1.1mm} \fmfset{dot_len}{0.5mm}
			\fmfpen{0.25mm}
			\fmfleft{i}
			\fmfright{o}
			\fmf{phantom,tension=5}{i,v1}
			\fmf{wiggly,fore=black,tension=2.5}{v2,o}
			\fmf{wiggly,fore=black,left,tension=0.5}{v1,v2,v1}
			\fmf{wiggly,fore=black}{v1,v2}
			\fmffreeze
			\fmfforce{(1.1w,0.5h)}{o}
			\fmffreeze
			\fmfright{n,m}
			\fmf{wiggly,fore=black,tension=1}{o,n}
			\fmf{wiggly,fore=black,tension=1}{o,m}
			\fmfforce{(1.35w,0.9h)}{n}
			\fmfforce{(1.35w,0.1h)}{m}
		\end{fmfgraph}
	\end{fmffile}
\end{gathered}
\,\,\,\,\,\,, 
\!\! \!\!\!\!\!\!
\quad
\begin{gathered}
	\begin{fmffile}{wg4-tadpole-2pt-}
		\begin{fmfgraph}(25,25)
			\fmfset{dash_len}{1.2mm}
			\fmfset{wiggly_len}{1.1mm} \fmfset{dot_len}{0.5mm}
			\fmfpen{0.25mm}
			\fmftop{t}
			\fmfbottom{b}
			\fmfleft{l}
			\fmfright{r}
			\fmf{phantom,fore=black,tension=9}{t,u,v,b}
			\fmf{phantom,fore=black,tension=9}{l,s,x,r}
			\fmf{wiggly,fore=black,tension=.01,left}{u,v,u}
			\fmf{phantom,fore=black,tension=0.01}{s,x,s}
			\fmf{wiggly,fore=black,tension=1}{u,v}
			\fmf{wiggly,fore=black,tension=1}{x,r}
			\fmffreeze
			\fmfright{a,b}
			\fmf{wiggly,fore=black,tension=1}{r,a}
			\fmf{wiggly,fore=black,tension=1}{r,b}
			\fmffreeze
			\fmfforce{(1.2w,0.8h)}{a}
			\fmfforce{(1.2w,0.2h)}{b}
		\end{fmfgraph}
	\end{fmffile}
\end{gathered}\,\,\,\,\,
, 
\quad
\begin{gathered}
	\begin{fmffile}{wgkappa-4pt1looptadpole,1PR-}
		\begin{fmfgraph}(12,12)
			\fmfset{dash_len}{1.2mm}
			\fmfset{wiggly_len}{1.1mm} \fmfset{dot_len}{0.5mm}
			\fmfpen{0.25mm}
			\fmfleft{i1,i2}
			\fmfright{o1,o2}
			\fmf{wiggly,fore=black}{i1,v,o2}
			\fmf{wiggly,fore=black}{i2,v,o1}
			\fmffreeze
			\fmfright{x}
			\fmf{wiggly,fore=black}{x,v}
			\fmf{wiggly,fore=black,tension=0.7}{x,x}
			\fmfforce{(1.2w,0.5h)}{x}
		\end{fmfgraph}
	\end{fmffile}
\end{gathered}\,\,\,
\,\,\,\,\,\,\,, \!\!\!\!
\quad
\begin{gathered}
	\begin{fmffile}{wgkappa-2pt1looptadpole-1PR-}
		\begin{fmfgraph}(19,19)
			\fmfset{dash_len}{1.2mm}
			\fmfset{wiggly_len}{1.1mm} \fmfset{dot_len}{0.5mm}
			\fmfpen{0.25mm}
			\fmftop{t1,t2,t3}
			\fmfbottom{b1,b2,b3}
			\fmf{phantom}{t1,v1,b1}
			\fmf{phantom}{t2,v2,b2}
			\fmf{phantom}{t3,v3,b3}
			\fmf{wiggly,fore=black,right,tension=1}{v1,v2,v1}
			\fmf{wiggly,fore=black,right,tension=1}{v2,v3,v2}
			\fmf{wiggly,fore=black,tension=1}{v2,b2}
			\fmffreeze
			\fmfbottom{x,z}
			\fmf{wiggly,fore=black,tension=1}{b2,x}
			\fmf{wiggly,fore=black,tension=1}{b2,z}
			\fmfforce{(0.25w,-.18h)}{x}
			\fmfforce{(0.75w,-.18h)}{z}
		\end{fmfgraph}
	\end{fmffile}
\end{gathered}\!,
\dots 
$$
\end{itemize}
\end{itemize}

We are now ready to discuss how to cancel all cephalopod Feynman diagrams in any scalar field theory of the form (\ref{eq:Notation}). We will not discuss all details of the computation in this brief note, and rather refer the reader to the article \cite{EllisMavromatosSkliros15} where these results are studied in great detail. We start with the following conjecture:
\begin{itemize}
\item \emph{All cephalopods can be cancelled by an appropriate choice of local counter terms} $\delta g_N$, $\delta Z$
\end{itemize}
Given that there generically will also be additional Feynman diagrams (other than cephalopods) that will be naively divergent, the counter terms $\delta g_N$, $\delta Z$ will need to absorb all such contributions. 
This means that we can orthogonally decompose these,
\begin{equation}\label{eq:deltagN decom}
\delta g_N\dfn \delta_Rg_N+\delta_Sg_N,\qquad {\rm for}\qquad N=0,1,2,\dots;\qquad \delta Z = \delta_RZ+\delta_SZ,
\end{equation}
with $\delta_Sg_N$, $\delta_SZ$ absorbing all cephalopods (independently of whether or not they are naively infinite) and $\delta_Rg_N$, $\delta_RZ$, absorbing all remaining divergences. The counter term $\delta_R Z$ is fixed (e.g.~by requiring the quantum effective action have canonical kinetic term) after all other divergences have been cancelled. In the absence of derivative interactions $\delta_SZ=0$.

The result (of a fairly long but straightforward computation \cite{EllisMavromatosSkliros15}) is rather simple: it turns out that one can cancel all cephalopod Feynman diagrams by generalising the notion of `normal ordering' of an operator, $\mathcal{O}\rightarrow\,\, \,:\!\mathcal{O}\!:$, to what we will refer to as `complete normal ordering', $\mathcal{O}\rightarrow *\mathcal{O}*$, and applying complete normal ordering to the full bare Lagrangian:
\begin{equation}\label{eq:LB-CNO}
\mathcal{L}_B(\phi_B)\,\,\,\rightarrow\, \,\,*\mathcal{L}_B(\phi_B)*.
\end{equation}
Denoting by $W(J)$ the full generating function of renormalised connected Greens functions as defined above, `\emph{complete normal ordering}' of a generic operator $\mathcal{O}(\phi)\rightarrow *\mathcal{O}(\phi)*$ is defined by:
\begin{equation}\label{eq:CNO-dfn}
\boxed{* \,\mathcal{O}(\phi)\,*=\mathcal{O}(\delta_X)\,e^{-W(X)+W(0)+\int_zX(z)\phi(z)}\Big|_{X=0}}
\end{equation}
with the renormalisation condition $\delta_XW(X)|_{X=0}=G_1=0$. Notice the field appearing, $\phi$, is the renormalised field, see above. 
There also exists a unique inverse,
\begin{equation}\label{eq:mathcalF2}
\mathcal{O}(\phi)=\mathcal{O}(\delta_X)\, e^{W(X)-W(0)}* e^{\int_zX(z)\phi(z)}*\Big|_{X=0}
\end{equation}

The expectation value of complete normal ordered operators in the full interacting theory vanishes (in the absence of external sources), as can be shown in one line of algebra:
\begin{equation}\label{eq:<nOn>a}
\begin{aligned}
\big\langle\!* \mathcal{O}(\phi)\!*\! \big\rangle
&=\mathcal{O}(\delta_X)e^{W(J+X)-W(X)+W(0)}\Big|_{X=J=0}\\
&=e^{W(0)}\mathcal{O}(\delta_X)\cdot 1,
\end{aligned}
\end{equation}
which implies that if $\mathcal{O}(\delta_X)\cdot1=0$, then $\langle * \mathcal{O}(\phi)* \rangle=0$.

Now, to make the statement (\ref{eq:LB-CNO}) precise let us on account of (\ref{eq:deltagN decom}) define a new set of `reduced' bare couplings, $Z'=1+\delta_RZ$, $g_N^{B'}$, with:
$$
g_N^{B'}\dfn \mu^{(N-2)\epsilon}(g_N+\delta_R g_N)Z^{-N/2},\qquad g_N^{B}\dfn \mu^{(N-2)\epsilon}(g_N+\delta_R g_N+\delta_S g_N)Z^{-N/2}.
$$
Of course, $g_N^{B}$ are the full bare couplings, so that $\frac{dg_N^B}{d\mu}=0$, and these should be used to derive beta functions, $\beta_N\dfn \mu\frac{dg_N}{d\mu}$, RG flows, $g_N(\mu)$, etc. The counter terms, $\delta_Rg_N$, $\delta_RZ$, for the remaining diagrams are fixed by standard methods \cite{Collins}. Then, the precise statement that relates complete normal ordering to the choice of counter terms in (\ref{eq:Notation}) is:
\begin{equation}\label{eq:CNO1}
\begin{aligned}
\mathcal{L}_B(g_1^B,g_2^B,\dots;Z)=*\mathcal{L}_B(g_1^{B'},g_2^{B'},\dots;Z')*,
\end{aligned}
\end{equation}
and this can be used to read off the counter terms $\delta_Sg_N$, $\delta_SZ$.\footnote{Needless to say, for this prescription to work the bare action must contain the counter terms generated by complete normal ordering. Whether this is the case in a given theory can be checked easily on account of the combinatorial interpretation of Sec.~\ref{sec:CI}.} Within perturbation theory, complete normal ordering typically reduces the number of Feynman diagrams by a factor of 2 or more. The counter terms $\delta_Sg_N$ generically depend on $W(J)$, which in turn the path integral is trying to define. And so we end up with an integro-differential equation for $W(J)$, that is most easily dealt with in perturbation theory. And it is within perturbation theory that these results have been checked \cite{EllisMavromatosSkliros15} (up to three loops, conjectured to hold to all loop orders). Given this result, one can in effect also derive $W(J)$ with usual methods and simply drop all (1PI and 1PR) cephalopod diagrams from the final answer. This is all one needs, as this $W(J)$ also leads directly to the required counter terms that accomplish this cancellation. Details are provided in \cite{EllisMavromatosSkliros15}.

%%%%%%%%%%%%%%%%%%%%%%%%%%%%%%%%%%%%%%%%%%%%%%%%%%%%%%%%%%%%%%%%%%%%%%%%%%%%%
\section{Combinatorial Interpretation}\label{sec:CI}
Complete normal ordering has a useful combinatorial interpretation that is best exhibited by means of a few examples. For instance, the complete normal ordering of a $\phi^N$ term is given by a sum of partitions of $N$ indistinguishable elements, the number of such partitions being given by Bell's number, $B_N$. For $N=4$ there are $B_4=15$ partitions:
\begin{spreadlines}{0.6\baselineskip}
\begin{equation*}
\begin{aligned}
&
\bigg\{
\begin{array}{l}
\begin{gathered}
	\begin{fmffile}{4pt-self-0-}
		\begin{fmfgraph}(15,15)
			\fmfset{dash_len}{1.2mm}
			\fmfset{wiggly_len}{1.1mm} \fmfset{dot_len}{0.5mm}
			\fmfpen{0.25mm}
			\fmfsurroundn{i}{4}
			\fmfvn{decor.shape=circle,decor.filled=full, decor.size=2.5thin}{i}{4}
		\end{fmfgraph}
	\end{fmffile}
\end{gathered}
\phantom{\Bigg\{}\end{array}\!\!\!\!\!\!\bigg\}
\quad \Leftrightarrow\quad \phi^4
\\
&
%%%%%%%%%%%%%%%%%%%%%%
\left\{
\begin{array}{l}
\begin{gathered}
	\begin{fmffile}{4pt-self-1-}
		\begin{fmfgraph}(15,15)
			\fmfset{dash_len}{1.2mm}
			\fmfset{wiggly_len}{1.1mm} \fmfset{dot_len}{0.5mm}
			\fmfpen{0.25mm}
			\fmfsurroundn{i}{4}
			\fmfvn{decor.shape=circle,decor.filled=full, decor.size=2.5thin}{i}{4}
			\fmf{plain,fore=black}{i1,i2}
		\end{fmfgraph}
	\end{fmffile}
\end{gathered}\hspace{0.7cm}
\begin{gathered}
	\begin{fmffile}{4pt-self-2-}
		\begin{fmfgraph}(15,15)
			\fmfset{dash_len}{1.2mm}
			\fmfset{wiggly_len}{1.1mm} \fmfset{dot_len}{0.5mm}
			\fmfpen{0.25mm}
			\fmfsurroundn{i}{4}
			\fmfvn{decor.shape=circle,decor.filled=full, decor.size=2.5thin}{i}{4}
			\fmf{plain,fore=black}{i2,i3}
		\end{fmfgraph}
	\end{fmffile}
\end{gathered}\hspace{0.7cm}
\begin{gathered}
	\begin{fmffile}{4pt-self-3-}
		\begin{fmfgraph}(15,15)
			\fmfset{dash_len}{1.2mm}
			\fmfset{wiggly_len}{1.1mm} \fmfset{dot_len}{0.5mm}
			\fmfpen{0.25mm}
			\fmfsurroundn{i}{4}
			\fmfvn{decor.shape=circle,decor.filled=full, decor.size=2.5thin}{i}{4}
			\fmf{plain,fore=black}{i3,i4}
		\end{fmfgraph}
	\end{fmffile}
\end{gathered}\hspace{0.7cm}
\begin{gathered}
	\begin{fmffile}{4pt-self-4-}
		\begin{fmfgraph}(15,15)
			\fmfset{dash_len}{1.2mm}
			\fmfset{wiggly_len}{1.1mm} \fmfset{dot_len}{0.5mm}
			\fmfpen{0.25mm}
			\fmfsurroundn{i}{4}
			\fmfvn{decor.shape=circle,decor.filled=full, decor.size=2.5thin}{i}{4}
			\fmf{plain,fore=black}{i4,i1}
		\end{fmfgraph}
	\end{fmffile}
\end{gathered}\phantom{\Bigg\{}\\
\begin{gathered}
	\begin{fmffile}{4pt-self-6c-}
		\begin{fmfgraph}(15,15)
			\fmfset{dash_len}{1.2mm}
			\fmfset{wiggly_len}{1.1mm} \fmfset{dot_len}{0.5mm}
			\fmfpen{0.25mm}
			\fmfsurroundn{i}{4}
			\fmfvn{decor.shape=circle,decor.filled=full, decor.size=2.5thin}{i}{4}
			\fmf{plain,fore=black}{i1,i3}
		\end{fmfgraph}
	\end{fmffile}
\end{gathered}\hspace{0.7cm}
\begin{gathered}
	\begin{fmffile}{4pt-self-7c-}
		\begin{fmfgraph}(15,15)
			\fmfset{dash_len}{1.2mm}
			\fmfset{wiggly_len}{1.1mm} \fmfset{dot_len}{0.5mm}
			\fmfpen{0.25mm}
			\fmfsurroundn{i}{4}
			\fmfvn{decor.shape=circle,decor.filled=full, decor.size=2.5thin}{i}{4}
			\fmf{plain,fore=black}{i2,i4}
		\end{fmfgraph}
	\end{fmffile}
\end{gathered}
\phantom{\Bigg\{}\end{array}\!\!\!\!\!\right\}
\quad \Leftrightarrow\quad -6\,G_2\phi^2
\\
&
%%%%%%%%%%%%%%%%%%%%%%
\left\{
\begin{array}{l}
\begin{gathered}
	\begin{fmffile}{4pt-self-3p-1-}
		\begin{fmfgraph}(15,15)
			\fmfset{dash_len}{1.2mm}
			\fmfset{wiggly_len}{1.1mm} \fmfset{dot_len}{0.5mm}
			\fmfpen{0.25mm}
			\fmfsurroundn{i}{4}
			\fmfvn{decor.shape=circle,decor.filled=full, decor.size=2.5thin}{i}{4}
			\fmf{plain,fore=black}{i1,i2,i3,i1}
		\end{fmfgraph}
	\end{fmffile}
\end{gathered}\hspace{0.7cm}
\begin{gathered}
	\begin{fmffile}{4pt-self-3p-2-}
		\begin{fmfgraph}(15,15)
			\fmfset{dash_len}{1.2mm}
			\fmfset{wiggly_len}{1.1mm} \fmfset{dot_len}{0.5mm}
			\fmfpen{0.25mm}
			\fmfsurroundn{i}{4}
			\fmfvn{decor.shape=circle,decor.filled=full, decor.size=2.5thin}{i}{4}
			\fmf{plain,fore=black}{i2,i3,i4,i2}
		\end{fmfgraph}
	\end{fmffile}
\end{gathered}\hspace{0.7cm}
\begin{gathered}
	\begin{fmffile}{4pt-self-3p-3-}
		\begin{fmfgraph}(15,15)
			\fmfset{dash_len}{1.2mm}
			\fmfset{wiggly_len}{1.1mm} \fmfset{dot_len}{0.5mm}
			\fmfpen{0.25mm}
			\fmfsurroundn{i}{4}
			\fmfvn{decor.shape=circle,decor.filled=full, decor.size=2.5thin}{i}{4}
			\fmf{plain,fore=black}{i3,i4,i1,i3}
		\end{fmfgraph}
	\end{fmffile}
\end{gathered}\hspace{0.7cm}
\begin{gathered}
	\begin{fmffile}{4pt-self-3p-4-}
		\begin{fmfgraph}(15,15)
			\fmfset{dash_len}{1.2mm}
			\fmfset{wiggly_len}{1.1mm} \fmfset{dot_len}{0.5mm}
			\fmfpen{0.25mm}
			\fmfsurroundn{i}{4}
			\fmfvn{decor.shape=circle,decor.filled=full, decor.size=2.5thin}{i}{4}
			\fmf{plain,fore=black}{i4,i1,i2,i4}
		\end{fmfgraph}
	\end{fmffile}
\end{gathered}
\phantom{\Bigg\{}\end{array}\!\!\!\!\!\right\}
\quad \Leftrightarrow\quad -4\,G_3\,\phi
\\
%%%%%%%%%%%%%%%%%%%%%%%%%
&
\left\{
\begin{array}{l}
\begin{gathered}
	\begin{fmffile}{4pt-self-22p-1-}
		\begin{fmfgraph}(15,15)
			\fmfset{dash_len}{1.2mm}
			\fmfset{wiggly_len}{1.1mm} \fmfset{dot_len}{0.5mm}
			\fmfpen{0.25mm}
			\fmfsurroundn{i}{4}
			\fmfvn{decor.shape=circle,decor.filled=full, decor.size=2.5thin}{i}{4}
			\fmf{plain,fore=black}{i1,i2}
			\fmf{plain,fore=black}{i3,i4}
		\end{fmfgraph}
	\end{fmffile}
\end{gathered}\hspace{0.7cm}
\begin{gathered}
	\begin{fmffile}{4pt-self-22p-2-}
		\begin{fmfgraph}(15,15)
			\fmfset{dash_len}{1.2mm}
			\fmfset{wiggly_len}{1.1mm} \fmfset{dot_len}{0.5mm}
			\fmfpen{0.25mm}
			\fmfsurroundn{i}{4}
			\fmfvn{decor.shape=circle,decor.filled=full, decor.size=2.5thin}{i}{4}
			\fmf{plain,fore=black}{i2,i3}
			\fmf{plain,fore=black}{i1,i4}
		\end{fmfgraph}
	\end{fmffile}
\end{gathered}\phantom{\Bigg\{}\\
\begin{gathered}
	\begin{fmffile}{4pt-self-22p-3-}
		\begin{fmfgraph}(15,15)
			\fmfset{dash_len}{1.2mm}
			\fmfset{wiggly_len}{1.1mm} \fmfset{dot_len}{0.5mm}
			\fmfpen{0.25mm}
			\fmfsurroundn{i}{4}
			\fmfvn{decor.shape=circle,decor.filled=full, decor.size=2.5thin}{i}{4}
			\fmf{plain,fore=black}{i1,i3}
			\fmf{plain,fore=black}{i2,i4}
		\end{fmfgraph}
	\end{fmffile}
\end{gathered}
\phantom{\Bigg\{}\end{array}\!\!\!\!\right\}
\quad \Leftrightarrow\quad 3\,G_2^2
\\
&
%%%%%%%%%%%%%%%%%%%%%%%%%%%
\bigg\{
\begin{array}{l}
\begin{gathered}
	\begin{fmffile}{4pt-self-4p-1-}
		\begin{fmfgraph}(15,15)
			\fmfset{dash_len}{1.2mm}
			\fmfset{wiggly_len}{1.1mm} \fmfset{dot_len}{0.5mm}
			\fmfpen{0.25mm}
			\fmfsurroundn{i}{4}
			\fmfvn{decor.shape=circle,decor.filled=full, decor.size=2.5thin}{i}{4}
			\fmf{plain,fore=black}{i1,i2,i3,i4,i1}
		\end{fmfgraph}
	\end{fmffile}
\end{gathered}
\phantom{\Bigg\{}\end{array}\!\!\!\!\!\bigg\}
\quad \Leftrightarrow\quad -G_4
\end{aligned}
\end{equation*}
\end{spreadlines}
The correspondence relating the diagrammatic representations, $\{\dots\}$, and operators (denoted by $\{\dots\}\Leftrightarrow f(\phi)$ above), is such that a collection of, say, $Q$ disconnected dots, `$\!\!\!\!\!\!\!\!\!\!
\begin{gathered}
	\begin{fmffile}{1pt-self-0-}
		\begin{fmfgraph}(15,15)
			\fmfset{dash_len}{1.2mm}
			\fmfset{wiggly_len}{1.1mm} \fmfset{dot_len}{0.5mm}
			\fmfpen{0.25mm}
			\fmfsurroundn{i}{1}
			\fmfvn{decor.shape=circle,decor.filled=full, decor.size=2.5thin}{i}{1}
		\end{fmfgraph}
	\end{fmffile}
\end{gathered}\,\,\,$', 
corresponds to $\phi^Q$, a collection of $M$ dots connected by a continuous line corresponds to the (negative of the) full renormalised $M$-point connected Greens function at coincident points,\footnote{In particular, $G_M\dfn \lim_{\{z_1,z_2,\dots\}\rightarrow z}G_M(z_1,\dots,z_M)$. These quantities are occasionally divergent and will require some regularisation procedure to make sense of them, but we wish to proceed in a scheme-independent manner for now. An explicit regularisation procedure is required in order to extract beta functions.} $-G_M$. Summing all partitions yields the complete normal ordered product:
\begin{equation}\label{eq:*phi4*}
\begin{aligned}
*\phi^4&*=\phi^4-6G_2\phi^2-4G_3\phi+3G_2^2-G_4.
\end{aligned}
\end{equation}
That is, the complete normal ordered product is obtained by subtracting all contractions using the full $N$-point connected Greens functions at coincident points.\footnote{From this viewpoint the term $+3G_2^2$ arises from the complete normal ordering of $\phi^2$, so that we could also write: $*\phi^4*=\phi^4-6G_2*\phi^2*-4G_3\phi-3G_2^2-G_4$, and note that $*\phi*=\phi$. } 
Generically, the map $\phi^N\rightarrow *\phi^N*$ (for $N=0,1,2,\dots$) is concisely expressed in terms of complete Bell polynomials,
\begin{equation}\label{eq:nphiNn}
*\phi^N*=B_N(\phi,-G_2,\dots,-G_N),
\end{equation}
the inverse, $*\phi^N*\rightarrow \phi^N$, being given by $\phi^N=\,*B_N(\phi,G_2,\dots,G_N)*$. 

Sometimes the monomial we want to complete normal order is not composed of indistinguishable elements; in this latter case the same pictorial representation as above still applies but (some of) the ``dots'' become distinguishable: for example, for a derivative interaction (relevant for non-linear sigma models), $\phi^2(\partial \phi)^2$, we may denote a derivative insertion, $\partial \phi$, by a circle, `$\!\!\!\!\!\!\!\!\!\!
\begin{gathered}
	\begin{fmffile}{1pt-self-0d-}
		\begin{fmfgraph}(15,15)
			\fmfset{dash_len}{1.2mm}
			\fmfset{wiggly_len}{1.1mm} \fmfset{dot_len}{0.5mm}
			\fmfpen{0.25mm}
			\fmfsurroundn{i}{1}
			\fmfvn{decor.shape=circle,decor.filled=empty, decor.size=2.5thin}{i}{1}
		\end{fmfgraph}
	\end{fmffile}
\end{gathered}\,\,\,$', and a $\phi$ insertion by a dot (as above), `$\!\!\!\!\!\!\!\!\!\!
\begin{gathered}
	\begin{fmffile}{1pt-self-0-}
		\begin{fmfgraph}(15,15)
			\fmfset{dash_len}{1.2mm}
			\fmfset{wiggly_len}{1.1mm} \fmfset{dot_len}{0.5mm}
			\fmfpen{0.25mm}
			\fmfsurroundn{i}{1}
			\fmfvn{decor.shape=circle,decor.filled=full, decor.size=2.5thin}{i}{1}
		\end{fmfgraph}
	\end{fmffile}
\end{gathered}\,\,\,$', and are led to the following results:
\begin{spreadlines}{0.6\baselineskip}
\begin{equation*}
\begin{aligned}
&
\bigg\{
\begin{array}{l}
\begin{gathered}
	\begin{fmffile}{4pt-self-der-0-}
		\begin{fmfgraph}(15,15)
			\fmfset{dash_len}{1.2mm}
			\fmfset{wiggly_len}{1.1mm} \fmfset{dot_len}{0.5mm}
			\fmfpen{0.25mm}
			\fmfsurroundn{i}{4}
			\fmfvn{decor.shape=circle,decor.filled=full, decor.size=2.5thin}{i}{4}
			\fmfv{decor.shape=circle,decor.filled=empty, decor.size=2.5thin}{i1}
			\fmfv{decor.shape=circle,decor.filled=empty, decor.size=2.5thin}{i2}
		\end{fmfgraph}
	\end{fmffile}
\end{gathered}
\phantom{\Bigg\{}\end{array}\!\!\!\!\!\!\bigg\}
\quad \Leftrightarrow\quad \phi^2(\partial\phi)^2
\\
&
%%%%%%%%%%%%%%%%%%%%%%
\left\{
\begin{array}{l}
\begin{gathered}
	\begin{fmffile}{4pt-self-der-1-}
		\begin{fmfgraph}(15,15)
			\fmfset{dash_len}{1.2mm}
			\fmfset{wiggly_len}{1.1mm} \fmfset{dot_len}{0.5mm}
			\fmfpen{0.25mm}
			\fmfsurroundn{i}{4}
			\fmfvn{decor.shape=circle,decor.filled=full, decor.size=2.5thin}{i}{4}
			\fmfv{decor.shape=circle,decor.filled=empty, decor.size=2.5thin}{i1}
			\fmfv{decor.shape=circle,decor.filled=empty, decor.size=2.5thin}{i2}
			\fmf{plain,fore=black}{i1,i2}
		\end{fmfgraph}
	\end{fmffile}
\end{gathered}\hspace{0.7cm}
\begin{gathered}
	\begin{fmffile}{4pt-self-der-2-}
		\begin{fmfgraph}(15,15)
			\fmfset{dash_len}{1.2mm}
			\fmfset{wiggly_len}{1.1mm} \fmfset{dot_len}{0.5mm}
			\fmfpen{0.25mm}
			\fmfsurroundn{i}{4}
			\fmfvn{decor.shape=circle,decor.filled=full, decor.size=2.5thin}{i}{4}
			\fmfv{decor.shape=circle,decor.filled=empty, decor.size=2.5thin}{i1}
			\fmfv{decor.shape=circle,decor.filled=empty, decor.size=2.5thin}{i2}
			\fmf{plain,fore=black}{i2,i3}
		\end{fmfgraph}
	\end{fmffile}
\end{gathered}\hspace{0.7cm}
\begin{gathered}
	\begin{fmffile}{4pt-self-der-3-}
		\begin{fmfgraph}(15,15)
			\fmfset{dash_len}{1.2mm}
			\fmfset{wiggly_len}{1.1mm} \fmfset{dot_len}{0.5mm}
			\fmfpen{0.25mm}
			\fmfsurroundn{i}{4}
			\fmfvn{decor.shape=circle,decor.filled=full, decor.size=2.5thin}{i}{4}
			\fmfv{decor.shape=circle,decor.filled=empty, decor.size=2.5thin}{i1}
			\fmfv{decor.shape=circle,decor.filled=empty, decor.size=2.5thin}{i2}
			\fmf{plain,fore=black}{i3,i4}
		\end{fmfgraph}
	\end{fmffile}
\end{gathered}\hspace{0.7cm}
\begin{gathered}
	\begin{fmffile}{4pt-self-der-4-}
		\begin{fmfgraph}(15,15)
			\fmfset{dash_len}{1.2mm}
			\fmfset{wiggly_len}{1.1mm} \fmfset{dot_len}{0.5mm}
			\fmfpen{0.25mm}
			\fmfsurroundn{i}{4}
			\fmfvn{decor.shape=circle,decor.filled=full, decor.size=2.5thin}{i}{4}
			\fmfv{decor.shape=circle,decor.filled=empty, decor.size=2.5thin}{i1}
			\fmfv{decor.shape=circle,decor.filled=empty, decor.size=2.5thin}{i2}
			\fmf{plain,fore=black}{i4,i1}
		\end{fmfgraph}
	\end{fmffile}
\end{gathered}\phantom{\Bigg\{}\\
\begin{gathered}
	\begin{fmffile}{4pt-self-der-6c-}
		\begin{fmfgraph}(15,15)
			\fmfset{dash_len}{1.2mm}
			\fmfset{wiggly_len}{1.1mm} \fmfset{dot_len}{0.5mm}
			\fmfpen{0.25mm}
			\fmfsurroundn{i}{4}
			\fmfvn{decor.shape=circle,decor.filled=full, decor.size=2.5thin}{i}{4}
			\fmfv{decor.shape=circle,decor.filled=empty, decor.size=2.5thin}{i1}
			\fmfv{decor.shape=circle,decor.filled=empty, decor.size=2.5thin}{i2}
			\fmf{plain,fore=black}{i1,i3}
		\end{fmfgraph}
	\end{fmffile}
\end{gathered}\hspace{0.7cm}
\begin{gathered}
	\begin{fmffile}{4pt-self-der-7c-}
		\begin{fmfgraph}(15,15)
			\fmfset{dash_len}{1.2mm}
			\fmfset{wiggly_len}{1.1mm} \fmfset{dot_len}{0.5mm}
			\fmfpen{0.25mm}
			\fmfsurroundn{i}{4}
			\fmfvn{decor.shape=circle,decor.filled=full, decor.size=2.5thin}{i}{4}
			\fmfv{decor.shape=circle,decor.filled=empty, decor.size=2.5thin}{i1}
			\fmfv{decor.shape=circle,decor.filled=empty, decor.size=2.5thin}{i2}
			\fmf{plain,fore=black}{i2,i4}
		\end{fmfgraph}
	\end{fmffile}
\end{gathered}
\phantom{\Bigg\{}\end{array}\!\!\!\!\!\right\}
\quad \Leftrightarrow\quad \,-(\partial^2G_2)\phi^2-4(\partial G_2)\phi\partial\phi-G_2(\partial\phi)^2
\\
&
%%%%%%%%%%%%%%%%%%%%%%
\left\{
\begin{array}{l}
\begin{gathered}
	\begin{fmffile}{4pt-self-der-3p-1-}
		\begin{fmfgraph}(15,15)
			\fmfset{dash_len}{1.2mm}
			\fmfset{wiggly_len}{1.1mm} \fmfset{dot_len}{0.5mm}
			\fmfpen{0.25mm}
			\fmfsurroundn{i}{4}
			\fmfvn{decor.shape=circle,decor.filled=full, decor.size=2.5thin}{i}{4}
			\fmfv{decor.shape=circle,decor.filled=empty, decor.size=2.5thin}{i1}
			\fmfv{decor.shape=circle,decor.filled=empty, decor.size=2.5thin}{i2}
			\fmf{plain,fore=black}{i1,i2,i3,i1}
		\end{fmfgraph}
	\end{fmffile}
\end{gathered}\hspace{0.7cm}
\begin{gathered}
	\begin{fmffile}{4pt-self-der-3p-2-}
		\begin{fmfgraph}(15,15)
			\fmfset{dash_len}{1.2mm}
			\fmfset{wiggly_len}{1.1mm} \fmfset{dot_len}{0.5mm}
			\fmfpen{0.25mm}
			\fmfsurroundn{i}{4}
			\fmfvn{decor.shape=circle,decor.filled=full, decor.size=2.5thin}{i}{4}
			\fmfv{decor.shape=circle,decor.filled=empty, decor.size=2.5thin}{i1}
			\fmfv{decor.shape=circle,decor.filled=empty, decor.size=2.5thin}{i2}
			\fmf{plain,fore=black}{i2,i3,i4,i2}
		\end{fmfgraph}
	\end{fmffile}
\end{gathered}\hspace{0.7cm}
\begin{gathered}
	\begin{fmffile}{4pt-self-der-3p-3-}
		\begin{fmfgraph}(15,15)
			\fmfset{dash_len}{1.2mm}
			\fmfset{wiggly_len}{1.1mm} \fmfset{dot_len}{0.5mm}
			\fmfpen{0.25mm}
			\fmfsurroundn{i}{4}
			\fmfvn{decor.shape=circle,decor.filled=full, decor.size=2.5thin}{i}{4}
			\fmfv{decor.shape=circle,decor.filled=empty, decor.size=2.5thin}{i1}
			\fmfv{decor.shape=circle,decor.filled=empty, decor.size=2.5thin}{i2}
			\fmf{plain,fore=black}{i3,i4,i1,i3}
		\end{fmfgraph}
	\end{fmffile}
\end{gathered}\hspace{0.7cm}
\begin{gathered}
	\begin{fmffile}{4pt-self-der-3p-4-}
		\begin{fmfgraph}(15,15)
			\fmfset{dash_len}{1.2mm}
			\fmfset{wiggly_len}{1.1mm} \fmfset{dot_len}{0.5mm}
			\fmfpen{0.25mm}
			\fmfsurroundn{i}{4}
			\fmfvn{decor.shape=circle,decor.filled=full, decor.size=2.5thin}{i}{4}
			\fmfv{decor.shape=circle,decor.filled=empty, decor.size=2.5thin}{i1}
			\fmfv{decor.shape=circle,decor.filled=empty, decor.size=2.5thin}{i2}
			\fmf{plain,fore=black}{i4,i1,i2,i4}
		\end{fmfgraph}
	\end{fmffile}
\end{gathered}
\phantom{\Bigg\{}\end{array}\!\!\!\!\!\!\right\}
\quad \Leftrightarrow\quad -2\,(\partial^2G_3)\,\phi-2\,(\partial G_3)\partial \phi
\\
%%%%%%%%%%%%%%%%%%%%%%%%%
&
\left\{
\begin{array}{l}
\begin{gathered}
	\begin{fmffile}{4pt-self-der-22p-1-}
		\begin{fmfgraph}(15,15)
			\fmfset{dash_len}{1.2mm}
			\fmfset{wiggly_len}{1.1mm} \fmfset{dot_len}{0.5mm}
			\fmfpen{0.25mm}
			\fmfsurroundn{i}{4}
			\fmfvn{decor.shape=circle,decor.filled=full, decor.size=2.5thin}{i}{4}
			\fmfv{decor.shape=circle,decor.filled=empty, decor.size=2.5thin}{i1}
			\fmfv{decor.shape=circle,decor.filled=empty, decor.size=2.5thin}{i2}
			\fmf{plain,fore=black}{i1,i2}
			\fmf{plain,fore=black}{i3,i4}
		\end{fmfgraph}
	\end{fmffile}
\end{gathered}\hspace{0.7cm}
\begin{gathered}
	\begin{fmffile}{4pt-self-der-22p-2-}
		\begin{fmfgraph}(15,15)
			\fmfset{dash_len}{1.2mm}
			\fmfset{wiggly_len}{1.1mm} \fmfset{dot_len}{0.5mm}
			\fmfpen{0.25mm}
			\fmfsurroundn{i}{4}
			\fmfvn{decor.shape=circle,decor.filled=full, decor.size=2.5thin}{i}{4}
			\fmfv{decor.shape=circle,decor.filled=empty, decor.size=2.5thin}{i1}
			\fmfv{decor.shape=circle,decor.filled=empty, decor.size=2.5thin}{i2}
			\fmf{plain,fore=black}{i2,i3}
			\fmf{plain,fore=black}{i1,i4}
		\end{fmfgraph}
	\end{fmffile}
\end{gathered}\phantom{\Bigg\{}\\
\begin{gathered}
	\begin{fmffile}{4pt-self-der-22p-3-}
		\begin{fmfgraph}(15,15)
			\fmfset{dash_len}{1.2mm}
			\fmfset{wiggly_len}{1.1mm} \fmfset{dot_len}{0.5mm}
			\fmfpen{0.25mm}
			\fmfsurroundn{i}{4}
			\fmfvn{decor.shape=circle,decor.filled=full, decor.size=2.5thin}{i}{4}
			\fmfv{decor.shape=circle,decor.filled=empty, decor.size=2.5thin}{i1}
			\fmfv{decor.shape=circle,decor.filled=empty, decor.size=2.5thin}{i2}
			\fmf{plain,fore=black}{i1,i3}
			\fmf{plain,fore=black}{i2,i4}
		\end{fmfgraph}
	\end{fmffile}
\end{gathered}
\phantom{\Bigg\{}\end{array}\!\!\!\!\!\right\}
\quad \Leftrightarrow\quad G_2\partial^2 G_2+2(\partial G_2)^2 
\\
&
%%%%%%%%%%%%%%%%%%%%%%%%%%%
\bigg\{
\begin{array}{l}
\begin{gathered}
	\begin{fmffile}{4pt-self-der-4p-1-}
		\begin{fmfgraph}(15,15)
			\fmfset{dash_len}{1.2mm}
			\fmfset{wiggly_len}{1.1mm} \fmfset{dot_len}{0.5mm}
			\fmfpen{0.25mm}
			\fmfsurroundn{i}{4}
			\fmfvn{decor.shape=circle,decor.filled=full, decor.size=2.5thin}{i}{4}
			\fmfv{decor.shape=circle,decor.filled=empty, decor.size=2.5thin}{i1}
			\fmfv{decor.shape=circle,decor.filled=empty, decor.size=2.5thin}{i2}
			\fmf{plain,fore=black}{i1,i2,i3,i4,i1}
		\end{fmfgraph}
	\end{fmffile}
\end{gathered}
\phantom{\Bigg\{}\end{array}\!\!\!\!\!\!\bigg\}
\quad \Leftrightarrow\quad -\partial^2G_4
\end{aligned}
\end{equation*}
\end{spreadlines}
and summing the partitions yields an expression for the complete normal ordered monomial,
\begin{equation}\label{eq:*phi2dphi2*}
\begin{aligned}
*\phi^2(\partial\phi)^2&*=\phi^2(\partial\phi)^2-(\partial^2G_2)\phi^2-4(\partial G_2)\phi\partial\phi-G_2(\partial\phi)^2\\
&\quad-2\,(\partial^2G_3)\,\phi-2\,(\partial G_3)\partial \phi+G_2\partial^2 G_2+2(\partial G_2)^2 -\partial^2G_4.
\end{aligned}
\end{equation}
The notation is such that $\partial^2G_N\dfn \lim_{\{z_j\}\rightarrow z}\partial_1\partial_{2}G_N(z_1,\dots,z_N)$, for all $N=2,3,\dots$, the Greens function is symmetric with respect to its arguments, and spacetime index contractions are implicit. Some of the terms in (\ref{eq:*phi2dphi2*}) are related by total derivatives, and the latter may or may not contribute depending on whether the couplings are local and whether the spacetime background is curved.

From the right-hand sides of (\ref{eq:*phi4*}) and (\ref{eq:*phi2dphi2*}) one can read off the terms that must be present in the bare Lagrangian in order to cancel all cephalopods associated with these interaction terms, in accordance with (\ref{eq:CNO1}).

Another instructive example is the case of an exponential interaction term (relevant for Liouville theory, see e.g.~\cite{Zams}), $e^{\,\mathfrak{g}\phi}$. To complete normal order this term we make use of (\ref{eq:nphiNn}) and a famous identity of complete Bell polynomials, leading to:
\begin{equation}\label{eq:nexpgphin}
*\exp\big(\mathfrak{g}\,\phi\big)*\,=\exp\bigg(\mathfrak{g}\,\phi-\sum_{N=2}^{\infty}\frac{1}{N!}\,\mathfrak{g}^NG_N\bigg).
\end{equation}
This is to be contrasted with free-field normal ordering, 
$
:\exp(\mathfrak{g}\,\phi)\!:\,\,=\exp(\mathfrak{g}\,\phi-\frac{1}{2}\,\mathfrak{g}^2\mathcal{G}).
$ 
Working with a complete normal ordered Liouville action leads to a renormalised generating function of connected Greens functions that is \cite{EllisMavromatosSkliros15} tadpole- (and more generally cephalopod-)free, thus overcoming the difficulties encountered in \cite{Zams}.

%%%%%%%%%%%%%%%%%%%%%%%%%%%%%%%%%%%%%%%%%%%%%%%%%%%%%%%%%%%%%%%%%%%%%%%%%%%%%
\section{Conclusions}
We have outlined how to cancel all tadpoles and more generally all cephalopod Feynman diagrams\footnote{This has been verified up to three loops within perturbation theory \cite{EllisMavromatosSkliros15} and conjectured to hold to all loop orders. In $n=2$ dimensions in the absence of derivative interactions the resulting Greens functions are UV finite, in $n>2$ further subtractions are needed. Derivative interactions are more subtle and two-derivative interactions are studied further in \cite{EllisMavromatosSkliros15}.} in generic scalar field theories on curved backgrounds. This was accomplished by introducing a generalisation of normal ordering that we call `complete normal ordering'.  In practical computations this means that one can drop all cephalopod Feynman diagrams from Greens functions of elementary fields, and the resulting Greens functions are all one needs in order to also write down the counter terms that are required to accomplish this cancellation. The detailed proofs of these conclusions and applications will be presented in the more detailed article \cite{EllisMavromatosSkliros15}, and applied to non-linear sigma models in \cite{EllisMavromatosSkliros15b}. We expect the notion of `complete normal ordering' to be particularly useful in various contexts, of which we mention two: in worldsheet and target space studies of quantum superstrings in non-trivial backgrounds (such as black holes or cosmological backgrounds); in the spontaneous breaking of supersymmetry in string theory  where typically the presence of massless tadpoles destabilises the vacuum and where there is a flourish of somewhat independent recent developments, see \cite{AngelantonjFlorakisTsulaia14}, \cite{Witten15}, and \cite{Sen15}, and references therein.

It is worth noting that complete normal ordering does not generically produce well-defined operators (i.e.~composite operators or normal products, see e.g.~\cite{Zimmermann73}). For this, certain consistency conditions need to be satisfied. For instance, if one tries to construct an operator product expansion (OPE), which can be derived from the definition of the map $\mathcal{O}\rightarrow *\mathcal{O}*$ and its inverse, $*\mathcal{O}*\rightarrow \mathcal{O}$, then for two (possibly non-local) operators, $* \mathcal{O}_1*$, $ *\mathcal{O}_2*$, one finds \cite{EllisMavromatosSkliros15}:
\begin{equation*}\label{eq:OPEv2}
\begin{aligned}
&* \mathcal{O}_1(\phi)*\,\,* \,\mathcal{O}_2(\phi)\,*=\mathcal{O}_1(\delta_{Y_1})\mathcal{O}_2(\delta_{Y_2})\, * \,e^{W(0)+W(Y_1+Y_2)-W(Y_1)-W(Y_2)+\int_z(Y_1+Y_2)\phi(z)}\!* \Big|_{Y_1,Y_2=0}
\end{aligned}
\end{equation*}
and applying this to, say, local exponential operators leads to:
\begin{equation*}\label{eq:OPEv2 exp}
\begin{aligned}
&* e^{\mathfrak{g}_1\phi(z)}\!*\,\,* \,e^{\mathfrak{g}_2\phi(w)}*=\exp\Big(\sum_{a,b=1}^{\infty}\frac{1}{a!b!}\,\mathfrak{g}_1^a\,\mathfrak{g}_2^b\,G_{a+b}(\underbrace{z,\dots,z}_{a},\underbrace{w,\dots,w}_{b})\Big)\,*e^{\mathfrak{g}_1\phi(z)+\mathfrak{g}_2\phi(w)}*,
\end{aligned}
\end{equation*}
which clearly contains Greens functions evaluated at coincident points, and this will \emph{generically} be divergent.\footnote{The author has benefited from discussions with Jo\~ao Penedones on this point.} Therefore, the operators $*\mathcal{O}(\phi)*$ cannot in general be identified with well-defined composite operators. For example, one might try to introduce a source term for $*\mathcal{O}(\phi)*$ in the action and add to it appropriate counter terms so as to produce finite Greens functions, and then one needs to check that the OPE gives a closed set of field operators. Note however that there do exist interacting quantum field theories for which the complete normal ordered operators are automatically well-defined composite operators, such as in the context of Liouville field theory. Here the basic primaries are the exponential fields, $V_{\alpha}=e^{2\alpha\phi}$, and Greens functions at coincident points are (up to a logarithmic divergence of the propagator) seemingly finite \cite{Zams}. 

Just like normal ordering, complete normal ordering is also not unique. One can replace the $N$-point Greens functions, $G_N$, (that appear in the defining relation $*\mathcal{O}(\phi)*$) by shifted Greens functions, $G'_N=G_N+\Delta_N$, and different choices of $\Delta_N$ give different complete normal ordering prescriptions. A specific prescription is required to cancel cephalopods completely (both the infinite and finite parts of these diagrams), namely $\Delta_N=0$. However, an alternative scheme is to choose $\Delta_N=-G_N$ for all $N\neq 2$, and $\Delta_2=-G_2+\mathcal{G}$, with $\mathcal{G}$ (as above) the free propagator. With this choice of scheme complete normal ordering reduces to the usual normal ordering, $*\mathcal{O}(\phi)*\rightarrow :\mathcal{O}(\phi):$, hence making it clear that normal ordering is a particular case of the more general definition of complete normal ordering in a particular scheme. 

%%%%%%%%%%%%%%%%%%%%%%%%%%%%%%%%%%%%%%%%%%%%%%%%%%%%%%%%%%%%%%%%%%%%%%%%%%%%%
\section*{Acknowledgements}
The author would like to thank John Ellis and Nikolaos E.~Mavromatos for collaboration on the work presented here and for feedback on the draft, and also Jo\~ao Penedones and Arkady Tseytlin for useful remarks and comments. 
This work was supported in part by the London Centre for Terauniverse Studies (LCTS), using funding from the European Research Council (ERC) via the Advanced Investigator Grant 267352. 

%\bibliographystyle{JHEP}
%\bibliography{spi-o}

\begin{thebibliography}{10}

\bibitem{Polchinski_v12}
J.~Polchinski, {\em String Theory. Vol. 1: An Introduction to the Bosonic
  String}.
\newblock Cambridge Univ. Pr., UK, 1998.; 
J.~Polchinski, {\em String Theory. Vol. 2: Superstring Theory and Beyond}.
\newblock Cambridge Univ. Pr., UK, 1998.

\bibitem{nlsm}
A.~A. Tseytlin, {\it {Sigma Model Weyl-Invariance Conditions and String
  Equations of Motion}},  {\em Nucl. Phys.} {\bf B294} (1987) 383; G.~M. Shore, {\it {A Local Renormalization Group Equation, Diffeomorphisms, and
  Conformal Invariance in $\sigma$ Models}},  {\em Nucl. Phys.} {\bf B286}
  (1987) 349; R.~R. Metsaev and A.~A. Tseytlin, {\it {Order $\alpha'$ (Two Loop) Equivalence
  of the String Equations of Motion and the Sigma Model Weyl Invariance
  Conditions: Dependence on the Dilaton and the Antisymmetric Tensor}},  {\em
  Nucl. Phys.} {\bf B293} (1987) 385--419; H.~Osborn, {\it {General Bosonic $\sigma$ Models and String Effective
  Actions}},  {\em Annals Phys.} {\bf 200} (1990) 1

\bibitem{HowePapadopoulosStelle88}
P.~S. Howe, G.~Papadopoulos, and K.~S. Stelle, {\it {The Background Field
  Method and the Nonlinear $\sigma$ Model}},  {\em Nucl. Phys.} {\bf B296}
  (1988) 26

\bibitem{RGstringloops}
W.~Fischler and L.~Susskind, {\it {Dilaton Tadpoles, String Condensates and
  Scale Invariance}},  {\em Phys.Lett.} {\bf B171} (1986) 383.; 
W.~Fischler and L.~Susskind, {\it {Dilaton Tadpoles, String Condensates and
  Scale Invariance. 2.}},  {\em Phys.Lett.} {\bf B173} (1986) 262.; 
J.~Polchinski, {\it {Factorization of Bosonic String Amplitudes}},  {\em Nucl.
  Phys.} {\bf B307} (1988) 61.; 
  C.~G. Callan, Jr., C.~Lovelace, C.~R. Nappi, and S.~A. Yost, {\it {String Loop
  Corrections to beta Functions}},  {\em Nucl. Phys.} {\bf B288} (1987) 525.; 
A.~A. Tseytlin, {\it {Renormalization group and string loops}},  {\em Int. J.
  Mod. Phys.} {\bf A5} (1990) 589--658.

\bibitem{CallanGan86}
J.~Callan, Curtis~G. and Z.~Gan, {\it {Vertex Operators in Background Fields}},
   {\em Nucl. Phys.} {\bf B272} (1986) 647; P.~Nelson, {\it Covariant Insertion of General Vertex Operators}, {\em Phys. Rev. Lett.} {\bf 62} (1989) 993--996

\bibitem{Zimmermann73}
W.~Zimmermann, {\it {Composite Operators in the Perturbation Theory of
  Renormalizable Interactions}},  {\em Annals Phys.} {\bf 77} (1973) 536--569.
  [Lect. Notes Phys.558,244(2000)]; 
W.~Zimmermann, {\it {Normal Products and the Short Distance Expansion in the
  Perturbation Theory of Renormalizable Interactions}},  {\em Annals Phys.}
  {\bf 77} (1973) 570--601. [Lect. Notes Phys.558,278(2000)]
  %STEINMANN, Perturbation Expansions in Axiomatic Field Theory, Theorem 2.1, p. 16,
%Springer-Verlag, Heidelberg, 1971.

\bibitem{Zams}
A.~B. Zamolodchikov and Al.~B. Zamolodchikov, {\it {Liouville Field Theory on a
  Pseudosphere}},  \href{http://arxiv.org/abs/hep-th/0101152}{{\tt
  hep-th/0101152}}

\bibitem{'tHooft73}
G.~'t~Hooft, {\it {Dimensional regularization and the renormalization group}},
  {\em Nucl.Phys.} {\bf B61} (1973) 455--468

\bibitem{EllisMavromatosSkliros15}
J.~Ellis, N.E.~Mavromatos, and D.~Skliros, {\it {Complete Normal Ordering, (to appear)}}

\bibitem{Collins}
John C.~Collins, (1984). {\it Renormalization}, { Cambridge: Cambridge University Press}

\bibitem{EllisMavromatosSkliros15b}
J.~Ellis, N.E.~Mavromatos, and D.~Skliros, {\it Complete Normal Ordering: Non-Linear Sigma Models {(to appear)}}
  
 % \bibitem{DineSeibergWitten87}
%M.~Dine, N.~Seiberg, and E.~Witten, {\it {Fayet-Iliopoulos Terms in String
%  Theory}},  {\em Nucl. Phys.} {\bf B289} (1987) 589.

\bibitem{AngelantonjFlorakisTsulaia14}
C.~Angelantonj, I.~Florakis, and M.~Tsulaia, {\it {Universality of Gauge
  Thresholds in Non-Supersymmetric Heterotic Vacua}},  {\em Phys. Lett.} {\bf
  B736} (2014) 365--370, [\href{http://arxiv.org/abs/1407.8023}{{\tt
  arXiv:1407.8023}}].

\bibitem{Witten15}
E.~Witten, {\it {The Super Period Matrix With Ramond Punctures}},  {\em J.
  Geom. Phys.} {\bf 92} (2015) 210--239,
  [\href{http://arxiv.org/abs/1501.02499}{{\tt arXiv:1501.02499}}]; 
E.~D'Hoker and D.~H. Phong, {\it {The Super Period Matrix with Ramond Punctures
  in the supergravity formulation}},  {\em Nucl. Phys.} {\bf B899} (2015)
  772--809, [\href{http://arxiv.org/abs/1501.02675}{{\tt arXiv:1501.02675}}].

\bibitem{Sen15}
A.~Sen, {\it {Supersymmetry Restoration in Superstring Perturbation Theory}},
  \href{http://arxiv.org/abs/1508.02481}{{\tt arXiv:1508.02481}}.


\end{thebibliography}

\end{document}